\documentclass{appolb}
\usepackage{graphicx}
\usepackage{ulem}
\usepackage{color}
\usepackage[numbers,sort&compress]{natbib}

\begin{document}
\title{A NOVEL EXPERIMENTAL SETUP FOR RARE EVENTS SELECTION AND ITS POTENTIAL APPLICATION TO SUPER HEAVY ELEMENTS SEARCH
}
\author{
    Z.~Majka$^a$\thanks{zbigniew.majka@uj.edu.pl}, R.~P\l{}aneta$^a$, Z.~Sosin$^a$, A.~Wieloch$^a$\thanks{andrzej.wieloch@uj.edu.pl},
    K.~Zelga$^a$, M.~Adamczyk$^a$, K.~Pelczar$^a$, M.~Barbui$^b$, S.~Wuenschel$^b$,
    K.~Hagel$^b$, X.~Cao$^b$, E-J.~Kim$^b$, J.~Natowitz$^b$, R.~Wada$^b$,
    H.~Zheng$^b$, G.~Giuliani$^b$, S.~Kowalski$^c$
\address{$^a$M. Smoluchowski Institute of Physics, Jagiellonian University, \L{}ojasiewicza 11, 30-348, Krak\'{o}w, Poland\\
         $^b$Cyclotron Institute, Texas A\&M University, USA\\
         $^c$University of Silesia, Katowice, Poland\\
         }
}
\maketitle

\begin{abstract}
The paper presents a novel  instrumentation for rare events selection which was tested in our research of short lived super heavy elements production and detection. The instrumentation includes an active catcher multi elements system and dedicated electronics. The active catcher located in the  forward hemisphere is composed of 63 scintillator detection modules. Reaction products of damped collisions between heavy ion projectiles and heavy target nuclei are implanted in the fast plastic scintillators of the active catcher modules. The acquisition system trigger delivered by logical branch of the electronics allows to record the reaction products which decay via the alpha particle emissions or spontaneous fission which take place between beam bursts. One microsecond wave form signal from FADCs contains information on heavy implanted nucleus as well as its decays. 
\end{abstract}

\section{Introduction}
\label{intro}
A frequent challenge for contemporary researchers in experimental physics is associated with the need to identify rare events out of the huge number of cases that are uninteresting. As examples of such investigations one can mention searches for the Higgs boson \cite{Aad:12:1} and neutrino-less double beta decay experiments \cite{Agos:17:1}. We are facing a similar problem in our searches of new super heavy elements (SHE). The question "How heavy can an atomic nucleus be?" is a fundamental problem in nuclear physics. The possible existence of island(s) of stable super heavy nuclei has been an inspiring problem in heavy ion physics for almost four decades \cite{Armb:00:1}. No stable or long life-times SHE (Z$>$103) has been found either in the natural environment of the Earth or in probes of meteorites or in cosmic rays. All have been produced artificially in complete fusion (CF) reactions between beam and target nuclei. Unfortunately, experimental studies have demonstrated that the cross section for SHE production in CF reactions is decreasing quite rapidly with the increasing atomic number, dropping for the synthesis of $^{277}_{112}$Cn to about 1 pb \cite{Hofm:96:1} and for a synthesis of element $^{294}_{118}$Og to about 0.5 pb \cite{Ogan:06:1}. Moreover, half-life times of the  SHEs are becoming  very short  decreasing to 0.7 ms for oganesson ($^{294}_{118}$Og). One of the possible explanations for these results is  that the newly produced elements were highly neutron deficient isotopes and they should in fact have quite short lifetimes.

From what was said above two basic conclusions can be drawn. Firstly, the CF experiments dedicated to super heavy nuclei synthesis require a large amount of the accelerator beam time, especially for nuclei with Z$>$118 (one can expect that the SHE production cross section in CF reactions will be in the region of tens of fb). As a consequence, a completely new generation of heavy ion sources is needed to supply the intensity of ion beams as high as 10$^{14}$-10$^{15}$ particles/s. This creates a serious limitation for the CF method being used so far. Secondly, available combinations of stable projectiles and targets cannot be used to produce neutron rich and longer lived SHEs in the predicted island of stability. 

In this context, another approach is urgently needed to achieve further progress in super heavy nuclei production. A promising possibility is to utilize multi nucleon transfer reactions occurring in collisions between heavy nuclei. Such reaction mechanisms have been already studied over  thirty years ago \cite{Hild:77:1,Gagg:80:1,Jung:78:1,Frei:79:1,Reid:79:1,Krat:86:1,Gagg:81:1,Scha:82:1}, however in both thin target and thick target irradiation experiments no new elements were observed. Although the cross-sections to produce SHE by multi nucleon transfer reactions occurring in collisions between heavy nuclei predicted theoretically are comparable with the cross-sections to the formation of SHE by a complete fusion method, the multi nucleon transfer processes in near barrier collisions of heavy and very heavy ions seem to be the only reaction mechanism (besides the multiple neutron capture process) allowing one to produce and explore neutron rich heavy nuclei including those located at the SHE island of stability \cite{Zagr:15:1}. Our research \cite{Donn:99:1,Barb:09:1:AW,Barb:10:1,Majk:14:1,Wiel:16:1}, which we are conducting
since year 1998, indicates that the collision process between heavy nuclei  leads to the creation of very heavy systems which disintegrate through the emission of highly energetic alpha particles which are our main signature of the very heavy systems formation. The arguments that we followed when undertaking and continuing this research are shortly summarized in the next section where we briefly outline the multi nucleon transfer concept of SHE creation.

Formation of SHE is a very rare event which should be selected out of the huge number of cases that are uninteresting. In section \ref{apparatus} we present a new concept and realization of  a detection system and dedicated electronics for registration of rare events in high intensity beam environment. The results of test measurements are shown in section \ref{testresults}. Suggestions to further developments of our experimental setup and conclusions are presented in section \ref{summary}.

\section{SHE production}
\label{production}
Our experimental research of  SHE production in collisions between very heavy nuclei was  initiated in the late 90s of the last century \cite{Donn:99:1}. A heavy projectile nucleus (e.g. $^{172}$Yb, $^{197}$Au) at a few MeV/nucleon incident energy goes into contact with a fissile target nucleus (e.g. $^{232}$Th, $^{238}$U). In the initial stage of the collision, a heavy projectile initiates deformation of the target nucleus  and nuclear interaction takes place between the objects  for a period long enough to transfer a large amount of mass to the projectile nucleus (e.g. by fusion of projectile nucleus with one of the target nucleus fission fragments). If such scenario takes place super heavy nucleus can be produced. 

Our early studies have indicated the possibility of forming in these reactions very heavy nuclei that emit high-energy alpha particles [15-16]. These results as well as other theoretical analyzes have motivated us to continue this research and to develop an innovative experimental approaches [17]. New stabilizing shell structures of very high Z nuclei as well as possible exotic shapes such as toroids and bubbles have been predicted \cite{Herm:79:1,Fler:83:1,Armb:99:1,Grei:99:1,Ogan:06:1,Ogan:15:1,Dech:99:1,Bend:01:1,Wong:73:1,Najm:15:1}.  Model calculations indicate existence of such stabilizing shell structures for nuclei from the islands of stability and predict that the fission barriers of these nuclei reduce the probability of spontaneous fssion \cite{Bend:13:1,Kire:12:1,Stas:13:1,Bara:15:1,Agbe:15:1,Agbe:17:1,Angh:17:1,Giul:17:1,Karp:12:1,Mart:12:1,Mark:16:1}. Thus the main modes of decay in and near these islands are predicted to be alpha and beta decay \cite{Kire:12:1,Stas:13:1,Karp:12:1,Mart:12:1,Mark:16:1}.  Predicted fission barriers and alpha decay energies rely upon model-dependent mass surface extrapolations \cite{Kire:12:1,Stas:13:1,Bara:15:1,Agbe:15:1,Agbe:17:1,Angh:17:1,Giul:17:1,Karp:12:1,Mart:12:1,Mark:16:1}. The predicted survival of heavy and super-heavy nuclei are extremely sensitive to details of these mass surface extrapolations and the location of closed shells. Uncertainties of 1 MeV in the fission barriers can lead to an order of magnitude change in the fission probabilities due to quantal effects of the barrier penetration \cite{Bara:15:1}. Uncertainties in level densities, temperature dependencies of fission barriers and details of the fission dynamics further complicate calculations of fission probabilities. While quantitative predictions vary widely, systematic theoretical studies indicate high survival probabilities of nuclei in and near the island of stability \cite{Kire:12:1,Stas:13:1,Bara:15:1,Angh:17:1,Giul:17:1,Karp:12:1,Mart:12:1,Mark:16:1}. Notably, recent microscopic fission model results indicate significant increases in fission survivability compared to those of statistical models employing the same fission barriers \cite{Zhu:17:1,Xia:11:1} and a strong increase in survivability is already evident in the experimental fusion cross section data for the heaviest elements \cite{Hami:15:1,Hofm:15:1,Utyo:16:1,Ogan:17:1}. Also some  calculations suggest that near the valley of stability, beta decay competes with alpha and spontaneous fission decay and that short-lifetime beta  minus  decay  will  be  dominant for the more neutron rich isotopes in that region \cite{Karp:12:1,Mart:12:1,Mark:16:1}. This raises the interesting possibility  that  the  production of neutron rich lower Z products can feed higher Z products through $\beta^{-}$ decay, increasing the effective production cross section for such higher Z products near the line of stability. 
Recent systematic efforts to explore the utility of multi-nucleon transfer reactions for production of new neutron-rich isotopes suggest that the experimental cross sections  for projectile (target) like fragment production exceed predicted cross sections  by 2-3 orders of magnitude \cite{Wels:17:1,Krat:15:1}. It is interesting to ask whether a similar trend exists for heavier elements. The production of alpha particle decaying heavy nuclei produced in massive transfer reaction between heavy nuclei has been explored in our recent research \cite{Wuen:15:1} using an in-beam detection array composed of YAP scintillators instead fast scintillators used in our work presented in this paper. Heavy nuclei with Z as high as 116, and perhaps higher, are being observed in these reactions what justify our innovative approach to the production of super heavy nuclei. Good experimental data are needed to guide future efforts in heavy element research.

\section {Experimental apparatus} 
\label{apparatus}
\begin{figure}
\begin{center}
\includegraphics[scale=0.043]{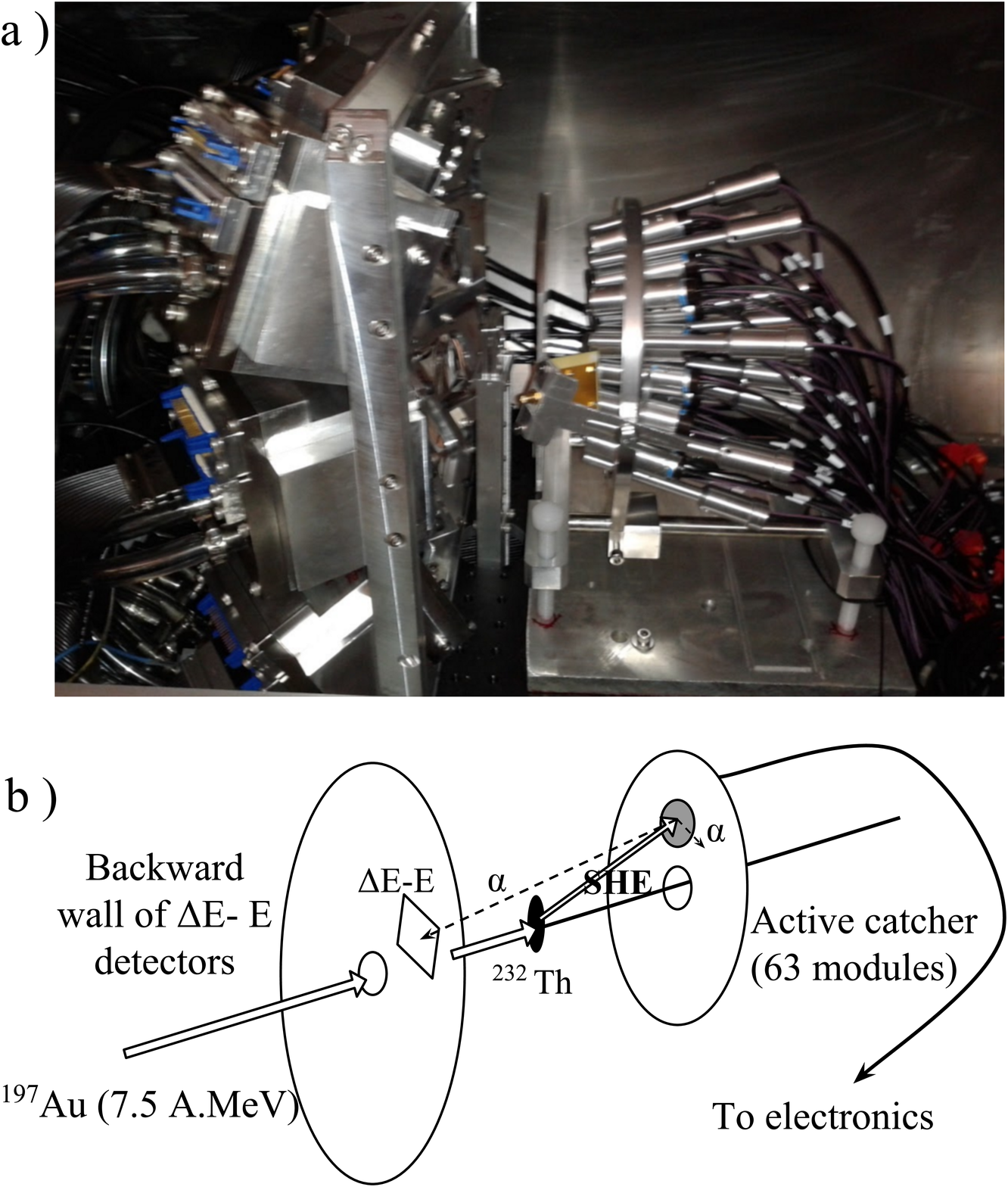}
\caption{The active catcher detection system (the right side  in the panel a) located behind the target (a bar in the middle of the panel a) and the backward wall of the gas – Si detectors (the left side in the panel a).  A schematic visualization of the detection setup (panel b).}
\label{fig:1} 
\end{center}
\end{figure}

The construction of the detection system used in the test measurement reported in this paper was based on experience collected during a decade of our experimental searches of SHEs.  A picture of the experimental setup is presented in Fig. \ref{fig:1}a and its schematic visualization is shown in Fig. \ref{fig:1}b. The detection system is composed of two separated units i.e. the forward hemisphere active catcher (AC) detection system composed of 63 scintillator detection modules and a set of ionization chambers equipped with 7 strip position sensitive Si detectors ($\Delta$E-E) placed at backward angles. We focus in this paper on the AC  detection system which allows to select candidates for  a short lived SHE production  out of large number of other uninteresting reaction products.  The reaction products of collisions between heavy projectiles and targets are deposited in the AC  modules and some of them which are  radioactive heavy nuclei  will decay by emission of alpha particles and/or by fission.  The active catcher detection system is only 10 cm from the target and can detect the creation of a radioactive nucleus with very short, even a few nano-seconds, half-lives.
The possibility of discovering the production of such short-lived SHEs was at the basis of the idea of the constructed apparatus.

\begin{figure}
\begin{center}
\includegraphics[scale=0.077]{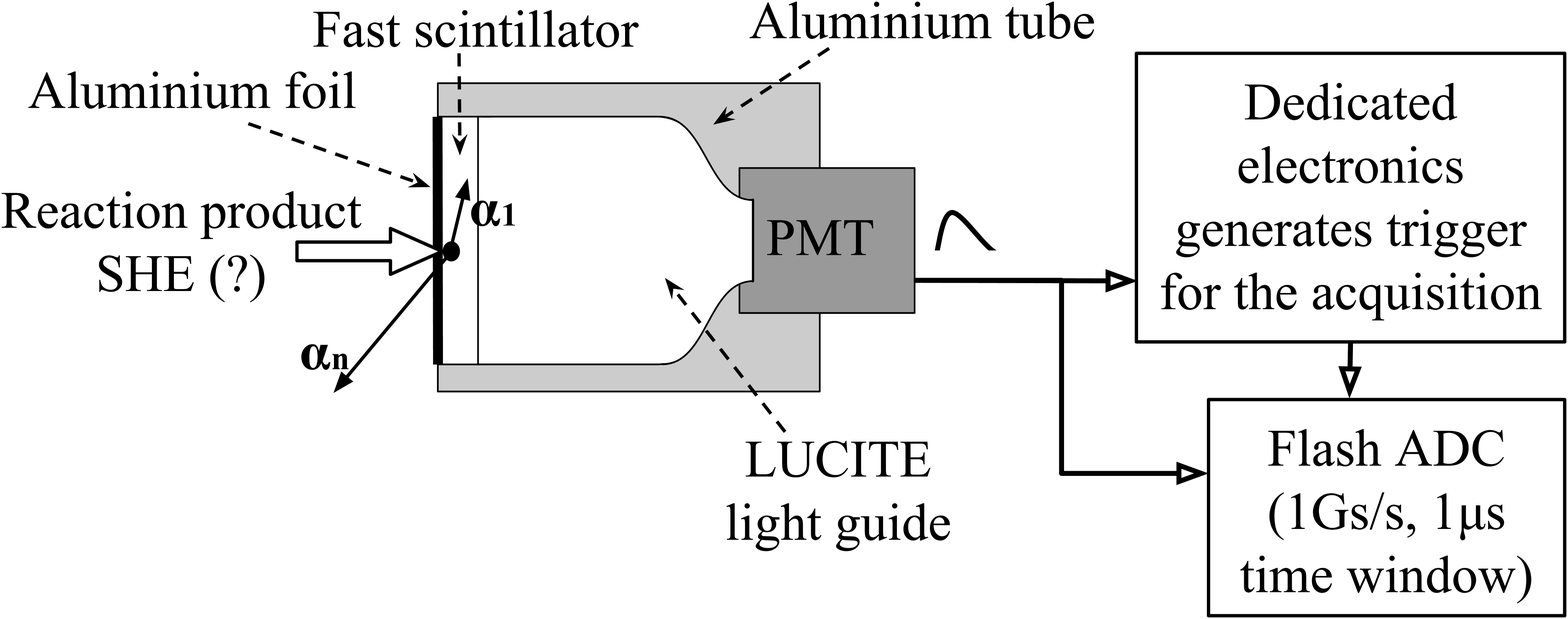}
\caption{Schematic drawing of the active catcher detection module. }
\label{fig:2} 
\end{center}
\end{figure}

The active catcher detection element presented in Fig.  \ref{fig:2} consists of fast plastic scintillator of 0.8 mm thickness, an aluminum cylinder with a cavity to accommodate a light guide and a photomultiplier tube (PMT). The light signals generated in the fast scintillator by the implanted reaction product and alpha particles and/or fission fragments emitted from the implanted heavy nucleus are converted by the PMT into electrical pulses which are processed by dedicated electronics. 

\begin{figure}
\begin{center}
\includegraphics[scale=0.05]{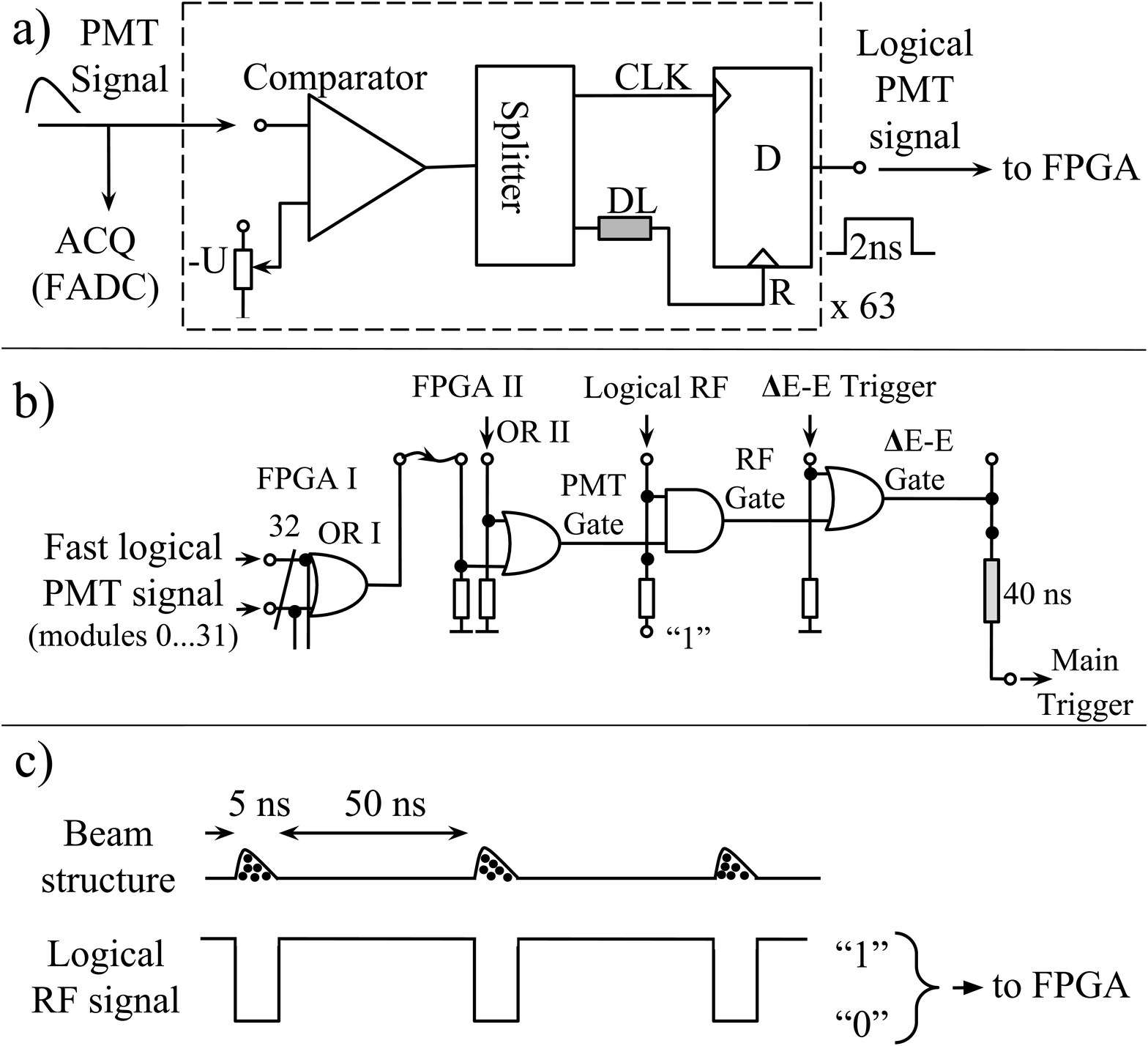}
\caption{A schematic drawing of the electronics. }
\label{fig:3} 
\end{center}
\end{figure}

The PMT signal from each detection module of the active catcher is split and sent into analog and digital logic branches of the electronics (see  Fig.  \ref{fig:3}a).
The main trigger produced by the logical branch of the electronics allows the recording of a signal wave form using  the CAEN FADC V1742 digitizer module. This module was set to a sampling rate of 1 Gs/s and 1024 points buffer. Therefore each event covers  a time window of 1 $\mu$s. In order to manage a very high signal rate caused by a high intensity of reaction products and to record information on the SHE candidate production, the main acquisition trigger is generated by logical electronics presented in  Fig.   \ref{fig:3}b.  For this experiment the beam structure of Texas A\&M University accelerator consisted of beam bursts of 5 ns width separated by   50 ns. The cyclotron RF signal is used to generate a logical veto to disable event recording during the beam burst (see Fig. \ref{fig:3}c). The fast plastic scintillator BC-418 prepared by Saint-Gobain Crystals, used in the active catcher module,  generates pulses of 0.5 ns rise time and 1.4 ns decay time. This scintillators are coupled to a small size  Hamamatsu   R9880U-110  photomultiplier (active window of 8 mm  diameter)  by a lucite light guide.  Each active catcher detection module has a very good time resolution (PMT pulse width is about 5 ns and the rise time is of the order of ns). 

The PMT signal of the logical branch is sent to a comparator which allows a computer controlled setting of a detection threshold and then a fast logical signal of 2 ns width is generated. The logical signals from all active catcher modules are sent into a logical OR of FPGA card. If the signal  from the logical OR of the FPGA card  (2 ns width) does not coincide with the beam burst logical signal  generated from the cyclotron RF (2 ns width) the main trigger is generated. The trigger signal caused by  decays between beam bursts of the reaction products implanted into the active catcher scintillator can occur as fast as a few ns after beam burst ions hit the target (time of flight of the reaction products on a distance of about 10 cm between the target and the active catcher detection module). The main trigger starts recording of the signal wave forms from all active catcher modules. The FADC acquisition time window of 1 $\mu$s is divided into 600 ns and 400 ns intervals which are located before and after the trigger signal time, respectively and the acquisition system  records all signals from the active catcher modules 600 ns before and 400 ns following time intervals with respect to the trigger signal time. 
\section{Test measurement results}
\label{testresults}
\begin{figure}[h]
\begin{center}
\includegraphics[scale=0.49]{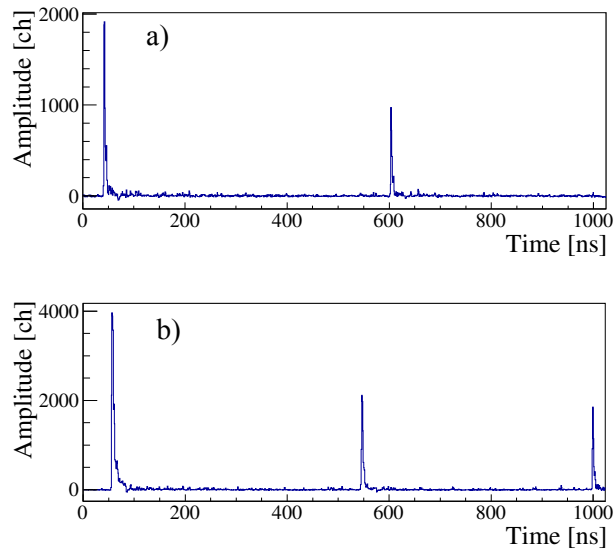}
\caption{The wave forms of two recorded events which might indicate on the production and observation of the SHEs. }
\label{fig:4} 
\end{center}
\end{figure}

Fig. \ref{fig:4} presents two examples of recorded events obtained in a summer 2015 experiment. A beam of $^{197}$Au (15-50 nA) at 7.5 A.MeV was delivered to the $^{232}$Th target of 12 mg/cm$^{2}$ thickness. Fig. \ref{fig:4}a shows the event when two signals were detected in only one of the active catcher modules. The pulse located at 602 ns is the triggering signal and represents decays of the implanted reaction product into the active catcher module scintillator which must occur between the beam bursts due to the triggering condition. The second peak at 42.5 ns may represent a signal from the deposition of the reaction product. The time distance between the two peaks is 559.5 ns. If this time interval is divided by 55 ns, i.e. the separation time between the beam bursts, the rest of division is 9.5 ns what is well beyond of the burst duration, Fig. 3c.  The peak at 42.5 ns precedes the  peak at 559.5 ns  and has much higher amplitude which may suggest that it originates from deposition of the reaction product produced during the beam burst. Moreover, we know that the peak at later time was generated by a particle emitted between beam bursts and we can conclude that this event can be a candidate for observation of implantation of the heavy reaction product which decays by the alpha particle emission after 517 ns plus a few ns needed by the heavy nucleus created in the target to travel about 10 cm distance to the active catcher module scintillator. 
We found about few tens similar cases among 1.5 million recorded events during our test measurements. The time interval between signals assigned as the implantation of the reaction product and the trigger signal assigned as the alpha particle emission from this reaction product covers the full range  of the FADC window i.e. 600 ns. 

In the collected data we also found a several of three peak events which may represent production and implantation of SHE into the active catcher module scintillator followed by two alpha particle emissions. An example of such a three pulse event is shown in Fig. \ref{fig:4}b.
 Both presented in this work as well as other collected cases for SHE candidates require more advanced analysis to confirm the production of very heavy nuclei in the massive transfer process.  Such analysis should allow for a more precise filtering of false signals and for more precise determination of the energy of particles which generate signals in active catcher detectors \cite{Wiel:17:1}. 


\begin{figure}[h]
\begin{center}
\includegraphics[scale=0.53]{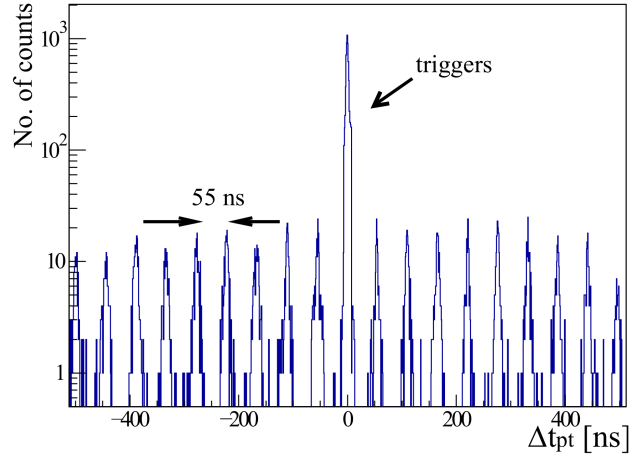}
\caption{Time spectrum of pulses locations  with respect to the trigger position.  }
\label{fig:5} 
\end{center}
\end{figure}

The stability and time resolution determination of the constructed electronics is visualized in Fig. \ref{fig:5} which shows a time spectrum of pulses’ positions recorded for all fired channels in one of the FADCs with respect to the trigger location. In order to accommodate sufficient statistics the triggers include also events associated with deposition in the triggering module beam burst reaction products. Observed regular structure of 55 ns period is a result of deposition in other detection modules reaction products associated with another beam bursts. The broadenings of the pulses’ positions are the result of around 5 ns beam burst width. Presented data proves that the electronics were stable and timing was determined with very high accuracy.

\section{Summary and conclusions}
\label{summary}

The article presents a new concept of the detection apparatus together with dedicated electronics for
registering rare events produced in nuclear reactions at high beam intensity. This concept has been
applied in our experimental searches for the production and detection of SHEs. The deposition of the
reaction product signal in the active catcher detection module as well as the signal of its decay via the
alpha particle emission or spontaneous fission which takes place between the beam bursts are
recorded. The FADC acquisition time window allows to record up to one microsecond separation between those signals. Preliminary results of the test measurements showed that the new concept and constructed apparatus allow for the selection and recording candidates for short-lived heavy nuclei
among other reaction products without overloading the acquisition system. The test run shows that
constructed detection system requires improvements to achieve better energy resolution and position determination of deposited reaction products. One
possibility is to use diamond detectors (2 mm by 2 mm active area), which have very good energy resolution (better than 10 keV) while
preserving their timing characteristics similar to that of the fast plastic scintillators. 

Authors are very indebted to the Cyclotron Institute crew for their great help and for  operation of the accelerator. This work is supported by the National Science Center in Poland, contract no. UMO-2012/04/A/ST2/
00082, by the U.S. Department of Energy under Grant No. DE-FG03-93ER40773 and by  the Robert A. Welch Foundation under Grant A0330.


\begin{thebibliography}{10}
\providecommand{\url}[1]{{#1}}
\providecommand{\urlprefix}{URL }
\expandafter\ifx\csname urlstyle\endcsname\relax
  \providecommand{\doi}[1]{DOI \discretionary{}{}{}#1}\else
  \providecommand{\doi}{DOI \discretionary{}{}{}\begingroup
  \urlstyle{rm}\Url}\fi

\bibitem{Aad:12:1}
G.~Aad et~al., Phys. Lett.
  \textbf{B716}, 1 (2012)

\bibitem{Agos:17:1}
M.~Agostini et~al., Nature \textbf{544}, 47 (2017)

\bibitem{Armb:00:1}
P.~Armbruster, Annu. Rev. Nucl. Part. S. \textbf{50}, 411 (2000)

\bibitem{Hofm:96:1}
S.~Hofmann et~al., Z. Phys. \textbf{A354}, 229 (1996)

\bibitem{Ogan:06:1}
Y.T. Oganessian et~al., Phys. Rev. \textbf{C74}, 044602 (2006)

\bibitem{Hild:77:1}
K.D. Hildenbrand et ~al., phys. Rev. Lett. \textbf{39}, 1065 (1977)

\bibitem{Gagg:80:1}
H.~G\"{a}ggeler et~al., Phys. Rev. Lett. \textbf{45}, 1824 (1980)

\bibitem{Jung:78:1}
H.~Jungclas et~al., Phys.
  Lett. \textbf{B79}, 58 (1978)

\bibitem{Frei:79:1}
H.~Freiesleben et~al., Z. Phys. \textbf{A 292}, 171 (1979)

\bibitem{Reid:79:1}
C.~Reidel, W.~Norenberg, Z. Physik \textbf{A290}, 385 (1979)

\bibitem{Krat:86:1}
J.V. Kratz et~al., Phys. Rev. \textbf{C33}, 504 (1986)

\bibitem{Gagg:81:1}
H.~G\"{a}ggeler et~al., Nucl. Inst. Meth. \textbf{A188}, 367 (1981)

\bibitem{Scha:82:1}
M.~Sch\"{a}del et~al., Phys. Rev. lett
  \textbf{48}, 852 (1982)

\bibitem{Zagr:15:1}
V.I~Zagrebaev, W. Greiner, Nucl. Phys. \textbf{A944},  257 (2015)
  
\bibitem{Donn:99:1}
T.W. O'{}Donnell et~al., Nucl. Inst. Meth. \textbf{A422}, 513 (1999)

\bibitem{Barb:09:1:AW}
M.~Barbui et~al., Int. Jour. Mod. Phys \textbf{E18}, 1036 (2009)

\bibitem{Barb:10:1}
M.~Barbui et~al., Nucl. Inst. Meth. \textbf{B268}, 20 (2010)

\bibitem{Majk:14:1}
Z.~Majka et~al., Acta Phys. Pol. \textbf{B45}, 279 (2014)

\bibitem{Wiel:16:1}
A.~Wieloch et~al., in \emph{EPJ Web of Conferences}
  (12TH INTERNATIONAL CONFERENCE ON NUCLEUS-NUCLEUS COLLISIONS 2015, 2016), p.
  01003

\bibitem{Herm:79:1}
G.~Hermann, Nature \textbf{280}, 543 (1979)

\bibitem{Fler:83:1}
G.N. Flerov, G.M. Ter-Akopian, Rep. Prog. Phys. \textbf{46}, 817 (1983)

\bibitem{Armb:99:1}
P.~Armbruster, Rep. Prog. Phys. \textbf{62}, 465 (1999)

\bibitem{Grei:99:1}
in \emph{Heavy Elements and Related New Phenomena}, ed. by W.~Greiner, R.K.
  Gupta (World Scientific, Singapore, London, 1999), p. 1 Part I

\bibitem{Ogan:15:1}
Y.T. Oganessian, V.K. Utyonkov, Rep. Prog. Phys. \textbf{78}, 036301 (2015)

\bibitem{Dech:99:1}
J.~Decharg'e et~al., Phys. Lett. \textbf{B451}, 275 (1999)

\bibitem{Bend:01:1}
M.~Bender et~al., Phys. Lett. \textbf{B515}, 42 (2001)

\bibitem{Wong:73:1}
C.Y. Wong, Ann. Phys.(N.Y.) \textbf{77}, 279 (1973)

\bibitem{Najm:15:1}
R.~Najman et~al., Phys. Rev. \textbf{C92}, 064614 (2015)

\bibitem{Bend:13:1}
M.~Bender, P.H. Heenen, Journal of Physics: Conference Series \textbf{420},
  012002 (2013)

\bibitem{Kire:12:1}
O.V. Kiren et~al., Rom. J. Phys. \textbf{57}, 1335
  (2012)

\bibitem{Stas:13:1}
A.~Staszczak et~al., Phys. Rev. \textbf{C87}, 024320 (2013)

\bibitem{Bara:15:1}
A.~Baran et~al., Nucl.
  Phys. \textbf{A944},  442 (2015)

\bibitem{Agbe:15:1}
S.E. Agbemava et~al., Phys. Rev.
  \textbf{C92}, 054310 (2015)

\bibitem{Agbe:17:1}
S.E. Agbemava et~al., Phys. Rev.
  \textbf{C95}, 054324 (2017)

\bibitem{Angh:17:1}
C.I. Anghel, I.~Silis, Phys. Rev. \textbf{C95}, 034611 (2017)

\bibitem{Giul:17:1}
S.A. Giuliani et~al., arXiv:1704.00554v1 [nucl-th]
  p. 3 Apr (2017)

\bibitem{Karp:12:1}
A.V. Karpov et~al., Int. J. Mod. Phys.
  \textbf{E21}, 1250013 (2012)

\bibitem{Mart:12:1}
Y.~Martinez-Palenzuela et~al., Bull. Russ. Acad. Sci.
  \textbf{76}, 1165 (2012)

\bibitem{Mark:16:1}
T.~Marketin et~al., Phys.Rev. \textbf{C93}, 025805
  (2016)

\bibitem{Zhu:17:1}
Y.~Zhu, J.C. Pei, arXiv:1709.04350v1 [nucl-th]  13 Sep (2017)

\bibitem{Xia:11:1}
C.J. Xia et~al., arXiv:1101.2725v1 [nucl-th]  14 Jan
  (2011)

\bibitem{Hami:15:1}
J.H. Hamilton et~al., J. Phys.: Conf. Ser. \textbf{580},
  012019 (2015)

\bibitem{Hofm:15:1}
S.~Hofmann, Y.T. Oganessian, Journal of Physics: Conference Series
  \textbf{580}, 012019 (2015)

\bibitem{Utyo:16:1}
V.~Utyonkov et~al., EPJ Web of
  Conferences \textbf{131}, 06003 (2016)

\bibitem{Ogan:17:1}
Y.~Oganessian, in \emph{FRYAA1 Proceedings of IPAC2017} (Copenhagen, Denmark,
  2017)

\bibitem{Wels:17:1}
T.~Welsh et~al., Phys. Lett. \textbf{B771}, 119 (2017)

\bibitem{Krat:15:1}
J.V. Kratz et~al., Nucl. Phys. \textbf{A944}, 117 (2015)

\bibitem{Wuen:15:1}
S. Wuenschel et~al., Phys. Rev. \textbf{C97}, 064602 (2017)

\bibitem{Wiel:17:1}
A.~Wieloch et~al., in preparation.

\end{thebibliography}
\end{document}